\documentclass[aps,prl,reprint,groupedaddress,amsmath,amssymb,showkeys]{revtex4-1}
\usepackage{graphicx}
\usepackage{dcolumn}
\usepackage{bm}
\usepackage{subcaption}
\begin{document}


\title{On the convergence of statistics in simulations of stationary turbulent flows}
\author {Yasaman Shirian}%
\email{yshirian@stanford.edu}
\author{Jeremy A. K. Horwitz}%
\thanks{Present~address: Lawrence Livermore National Laboratory, Livermore, CA~94550}
\email{horwitz3@llnl.gov}
\author{Ali Mani}%
\email{alimani@stanford.edu}
\affiliation{Center for Turbulence Research, Stanford University, Stanford, California 94305, USA}



\begin{abstract}
When reporting statistics from simulations of statistically stationary chaotic phenomenon, it is important to verify that the simulations are time-converged. This condition is connected with the statistical error or number of digits with which statistics can be reliably reported. In this work we consider homogeneous and isotropic turbulence as a model problem to investigate statistical convergence over finite simulation times. Specifically, we investigate the time integration requirements that allow meaningful reporting of the statistical error associated with finiteness of the temporal domain. We address two key questions: 1) What is the appropriate range of sampling frequency in large eddy time units? and 2) How long should a simulation be performed in terms of large eddy time so that the statistical error could be reliably reported. Our results indicate that proper sampling frequency is on the order of 10 large eddy time scale. More importantly, we find that reliable reporting of statistical errors requires simulation durations orders of magnitude longer than typically performed. Our observations of sampling frequency for homogeneous isotropic turbulence are also shown to hold in turbulent channel flow.
\end{abstract}
\keywords{Statistical error, Turbulence simulation}

\maketitle

\section{Introduction}

When analyzing turbulent flows, most quantities of interest are sought in terms of expected values or ensemble averages. For example, in aerodynamics often the mean lift and drag or mean pointwise skin friction are the target quantities that impact engineering design decisions. When a flow field is statistically stationary, these expected values, can be estimated from the time average over many instantaneous measurements. 
For an infinitely long  simulation, one could compute the ideal mean of the quantity of interest. Mathematically, if $q$ is a measured quantity  in a chaotic or turbulent flow, its mean can be expressed as 
\begin{equation}
\overline{q} = \textnormal{lim}_{T \rightarrow \infty} \quad  \frac{1}{T}\int_0^T q dt,
\label{equ:mean}
\end{equation}

\begin{figure}[htb!]
    \centering
    \includegraphics[width=0.4\textwidth]{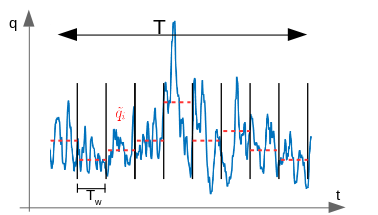}
    \caption{A schematic representing a quantity of interest $q$ from a simulation over finite time $T$. The simulation time is then divided into smaller windows of size $T_w$ with $\tilde{q}_i$ indicated by the dashed red lines representing the mean within each window.}
    \label{fig:qVSt}
\end{figure}

Where $T$ is the size of the time interval over which the measurement had taken place. Unlike experiments, which typically can be performed for sufficiently long times to obtain a fairly converged mean, numerical simulations of turbulent flows are considerably more susceptible to sampling error resulting from finite-time simulations. Therefore, it is advisable to report an estimated error of the mean along with the estimated mean when reporting statistics from simulation data. One way to estimate the error of the mean is to divide the finite simulation interval into $N$ smaller time windows with size $T_w$. As depicted in Figure \ref{fig:qVSt}, we can calculate the corresponding mean ($\tilde{q}_i$) for each window separately. The estimate of the mean in~(\ref{equ:mean}) can be reported as the average of means computed over all windows as:
\begin{equation}
\overline{q}_{est} = \frac{1}{N}\sum_i^N \tilde{q}_i.    
\end{equation}

The error quantifying the difference between $\overline{q}_{est}$ and $\overline{q}$ can itself be estimated using the standard error of the mean formula \cite{Ross}:
\begin{equation}
    \text{SEM}= \text{STD}(\overline{q}-\overline{q}_{est}) \simeq \frac{\text{STD}(\tilde{q}_i)}{\sqrt{N}},
    \label{equ:SEM}
\end{equation}

where the STD on the left hand side represents standard deviation over different realizations of $\overline{q}_{est}$ and the STD on the left is the standard deviation of the means from smaller windows. Equation (\ref{equ:SEM}) relies on the following key assumptions:
\begin{enumerate}
    \item Law of large numbers: $N$ should be large enough($N \gg 1$) so that the estimated $\text{STD}(\tilde{q}_i)$ be representative of the true STD. 
    \item Uncorrelated $\tilde{q}_i$ samples. In other words, the size of each window ($T_w$) should be large compared to the correlation time scale.
\end{enumerate}

For a fixed simulation interval, $T$, these two criteria conflict with each other. While the first criteria demands more windows for sampling, it forces smaller window sizes as $T_w$ is inversely proportional to $N$.
\begin{figure*}[htb!]
\begin{subfigure}{.49\textwidth}
\includegraphics[width=0.95\linewidth]{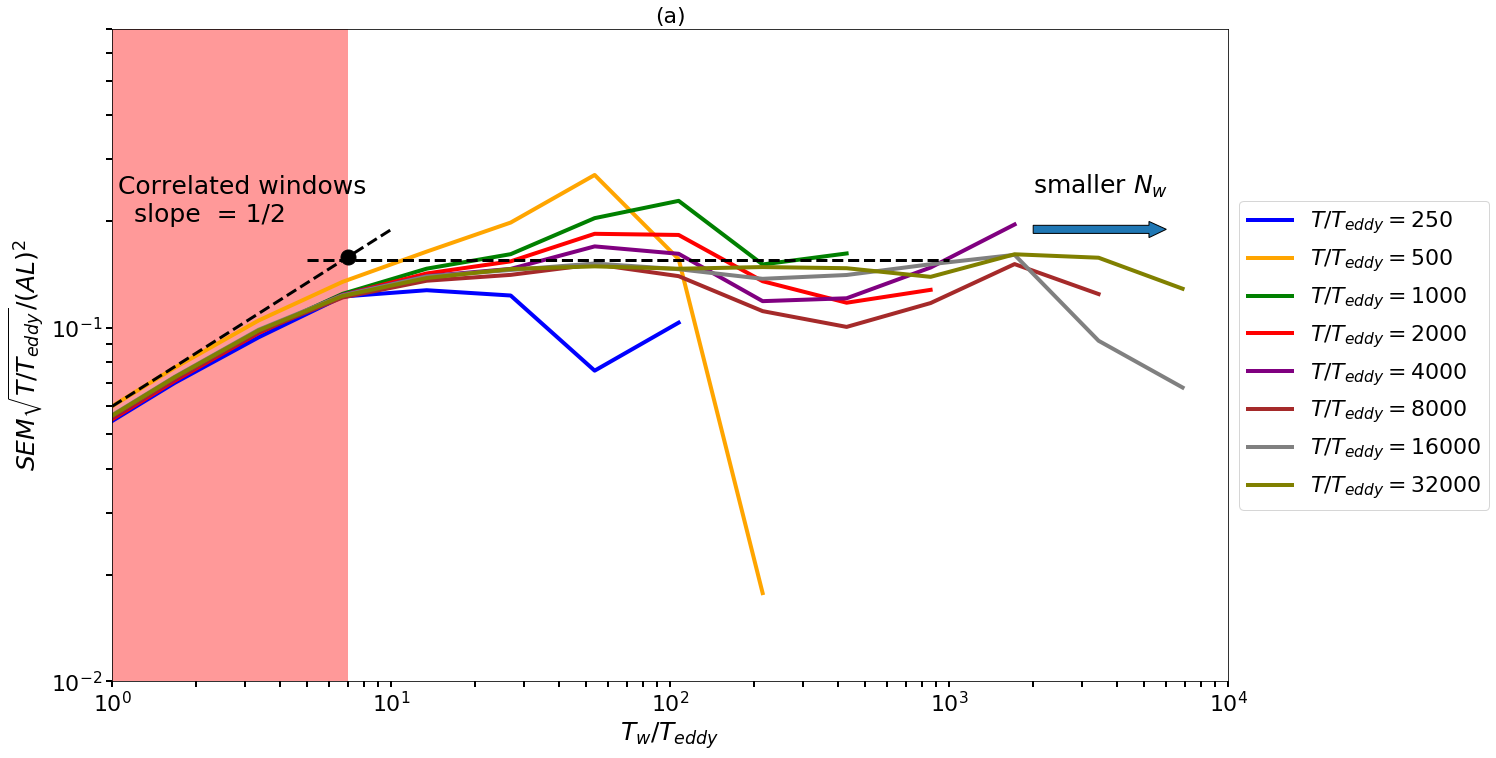}
\label{fig:Re26_uprime2}
\end{subfigure}
\begin{subfigure}{.49\textwidth}
\includegraphics[width=0.95\linewidth]{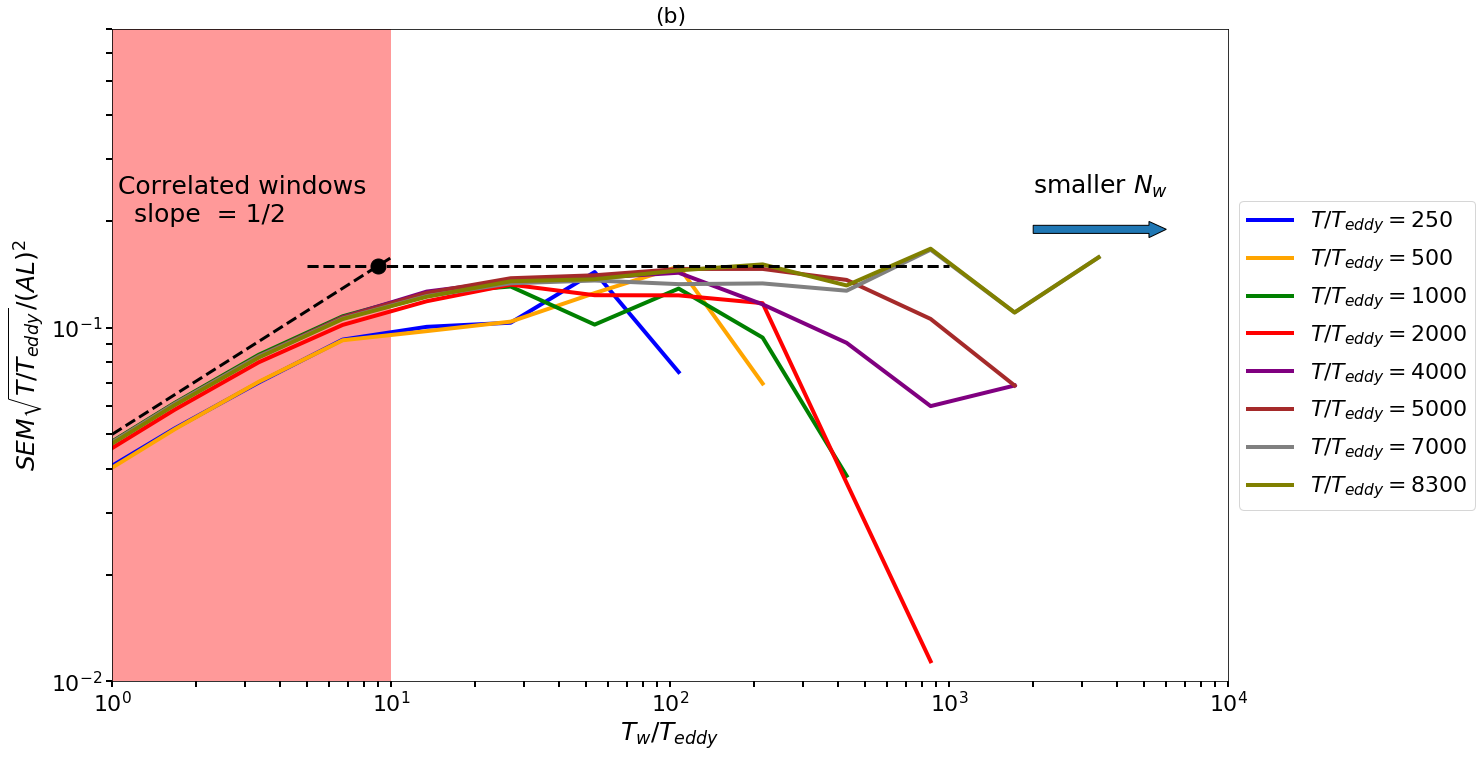}
\label{fig:Re40_uprime2}
\end{subfigure}

\begin{subfigure}{.49\textwidth}
\includegraphics[width=0.95\linewidth]{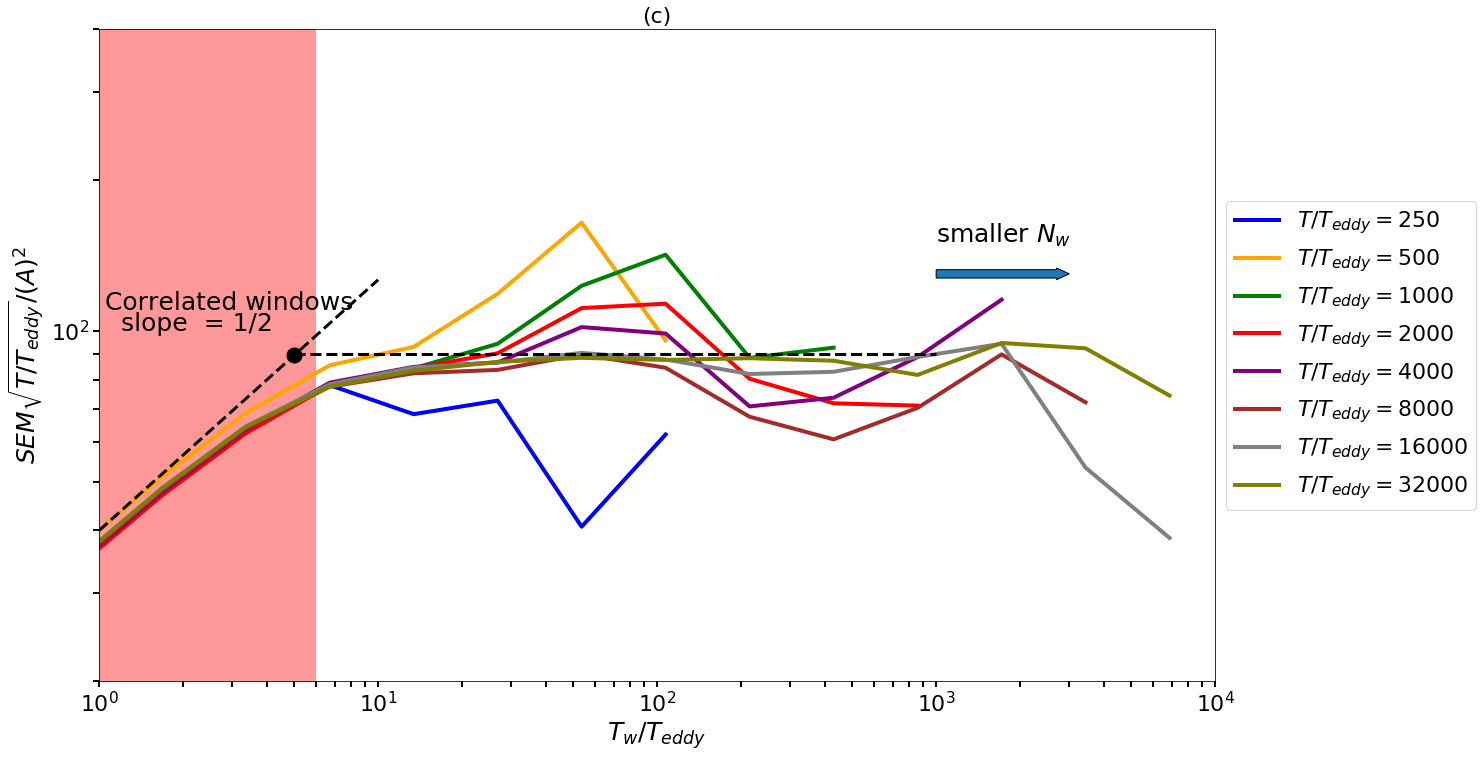}
\label{fig:Re26_SijSij}
\end{subfigure}
\begin{subfigure}{.49\textwidth}
\includegraphics[width=0.95\linewidth]{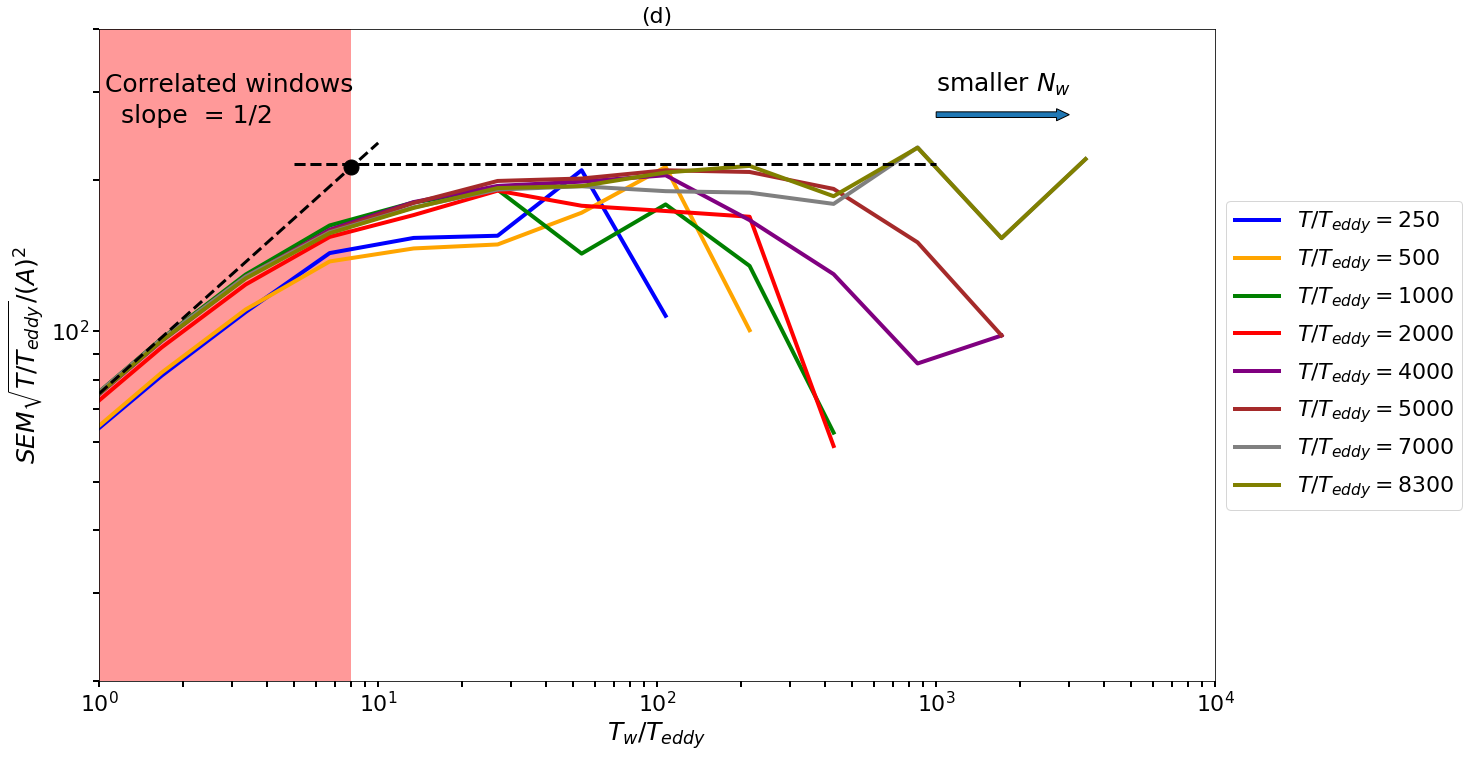}
\label{fig:Re40_SijSij}
\end{subfigure}
\caption{statistical error versus $\frac{T_w}{T_{eddy}}$ for a)$\overline{{u^{\prime}}^2}$ at $Re_\lambda=26$, b)$\overline{{u^{\prime}}^2}$ at Re=40, c)$\overline{S_{ij}S_{ij}}$ at Re=26 , d)$\overline{S_{ij}S_{ij}}$ at $Re_\lambda=40$. }
\label{fig:REMvsT}
\end{figure*}
One way to reconcile these trade-offs, is to perform longer simulations, thus ensuring T is large enough to satisfy both conditions. Our focus in this paper is to answer two simple yet important questions:
\begin{enumerate}
    \item What is the appropriate range of window size to ensure uncorrelated samples from different windows?
    \item What simulation time interval is needed to allow accurate estimation of the error in the reported mean?
    
\end{enumerate}

\section{Model Problem}

The model problem we considered is forced homogeneous isotropic turbulence providing statistically stationary data for our analysis.  Lundgren et al. (2003) \cite{Lundgren} proposed forcing the  Navier-Stokes equations with a linear function of velocity.
\begin{equation}
\frac{\partial u_{i}}{\partial t} + u_{j}\frac{\partial u_{i}}{\partial x_{j}} = -1/\rho \frac{\partial p}{\partial x_{i}} + \nu \frac{\partial ^2 u_{i}}{\partial x_{j} \partial x_{j}} + Au_i.
\label{equ:Navier-Stokes}
\end{equation}
For fully developed turbulence, the average input energy rate balances the dissipation rate
\begin{equation}
\epsilon = 2 \nu  \overline{{{S_{ij}}^{\prime}{S_{ij}}^{\prime}}}  = 3A u_\text{rms}^2.
\label{equ:TKE_forced}
\end{equation}
Rosales and Meneveau \cite {Meneveau}, estimated the forcing coefficient that needs to be specified to achieve a desired dissipation for a given periodic domain size $L$:
\begin{equation}
A = \frac{\epsilon ^{\frac{1}{3}}}{0.99 L^{\frac{2}{3}}}
\label{equ:A_Meneveau}
\end{equation}
Combining~(\ref{equ:TKE_forced}) and (\ref{equ:A_Meneveau}) one can eliminate $\epsilon$ to obtain an estimate of $A$ in terms of a desired $u_\text{rms}$. One should note however, that Equation~(\ref{equ:A_Meneveau}) is not exact and is only an approximate correlation. In this work we selected $A=0.2792$, targeting $u_\text{rms}\simeq1$ when $L=2\pi$. Table \ref{tab:simulation} lists the Reynolds numbers and the corresponding viscosity coefficients and mesh resolutions considered in our analysis. With these parameters set, we solved the incompressible forced Navier-Stokes equations using a second order spatial finite differences scheme and the fourth-order Runge Kutta scheme for time advancement\cite{Hadi} inside a triply periodic box of size $(2\pi)^3$.
\begin{table}[htb!]
\caption{Simulation parameters. }
    \label{tab:simulation}
    \begin{ruledtabular}
    \begin{tabular}{ccccc}
         $Re_{\lambda}$ & L & $N^3$ & $\nu$ & A \\
         \hline
         26 & $2\pi$ & $62^3$ & 0.0263 & 0.2792 \\
         \hline
         40 & $2\pi$ & $118^3$ & 0.0111 & 0.2792 \\
    \end{tabular}
    \end{ruledtabular}
\end{table}

We next select an appropriate quantity of interest, $q$, whose mean will be estimated from post processing of the simulation data. We then examine conditions pertaining to the accuracy of (\ref{equ:SEM}) in estimating the error of the mean, from which we extract criteria for appropriate sampling window and simulation time interval. 

As a starting point, we consider $q$ to be defined the square of the single component velocity magnitude, $q=u^2=u'^2$. For the purpose of this analysis we ignore knowledge of the approximate correlation in (\ref{equ:A_Meneveau}) and instead we solely rely on the simulation data to estimate any statistical quantity of interest. For this example, given homogeneity and isotropy of the flow, we compute $\tilde{q}_i$'s by taking average of $q$ over the box volume, all three components of the velocity field, and over time $T_w$ defining the size of each time window. Substitution of $\tilde{q}_i$'s in Equations (\ref{equ:mean}) and (\ref{equ:SEM}) provides numerical estimates of the mean and its error. Next, we assess the adequacy of the sampling condition as follows. 

In using Equation (\ref{equ:SEM}) one has some freedom in choosing the window size $T_w$. For a sufficiently large $T$ there must be a range of valid $T_w$'s that are all larger than the correlation time and all small enough to allow sufficiently large $N=T/T_w$ and thereby allowing both of the aforementioned validity conditions of (\ref{equ:SEM}) to be met. Furthermore, since the source data used in estimating the error of the mean does not change when changing $T_w$ the resulting estimate of the error from (\ref{equ:SEM}) should not depend on $T_w$ as long as both validity conditions are met. Therefore, for a fixed $T$ when plotting SEM versus $T_w$ one should identify an intermediate range of $T_w$ over which the plot is approximately a flat line. We can simply use this criteria to identify the minimum $T$ needed for reliable reporting of the statistical error of the mean. Additionally, the lower bound on $T_w$ marking the beginning of the flat zone allows identification of the minimum window size that would ensure uncorrelated $\tilde{q}_i$ samples.



\begin{figure*}
    \centering
    \includegraphics[width=0.6\textwidth]{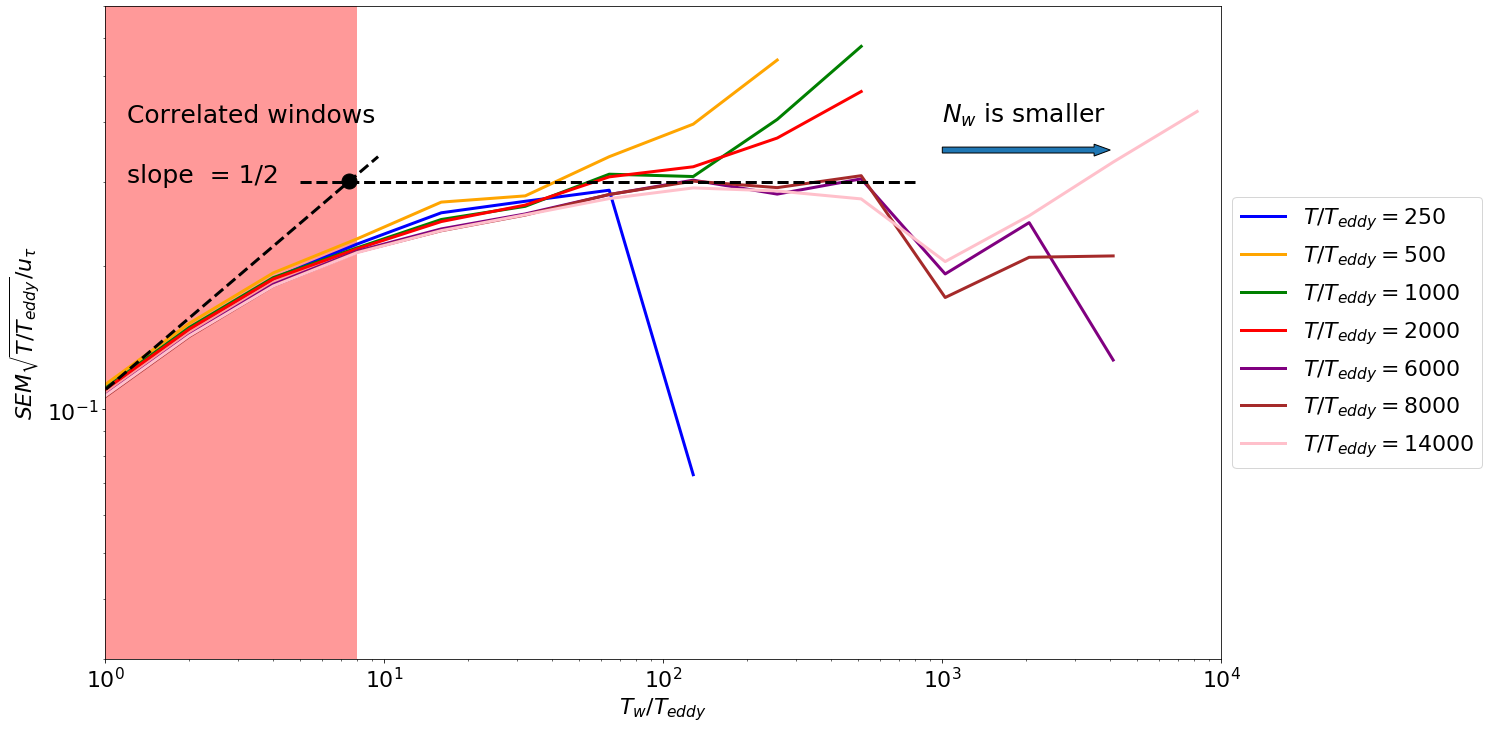}
    \caption{Statistical error of the mean stream-wise velocity $\overline{u}$ in Channel flow with $Re_\tau = 180$}
    \label{fig:Channel}
\end{figure*}


\section{Results}
Figure~\ref{fig:REMvsT}.a and ~\ref{fig:REMvsT}.b show the statistical error of the mean of $u^2$ versus $T_w$ obtained from the described procedure. Each curve shows the results for a different simulation time $T$. To allow better visual comparison, the SEM values are scaled by $\sqrt{T}$ allowing collapse of the curves in the flat zone, and all time scales are nondimensionalized by $T_\text{eddy}=1/A$. Additionally, in Figures~\ref{fig:REMvsT}.c and ~\ref{fig:REMvsT}.d same analysis is performed but with a different statistical quantity of interest, $q= S_{ij}S_{ij}$.


In all cases presented in Figure \ref{fig:REMvsT} three zones are noticeable.

The first zone marked by pink color corresponds to very small $T_w/T_{eddy}$; therefore data between windows are correlated. In this region, Equation (\ref{equ:SEM}) underpredicts the statistical error, since $N$ in its denominator over-estimates the number of truly independent samples. For this reason, the slope of dashed black line in the pink zone is $\frac{1}{2}$, illustrating underestimation of $SEM$ proportional to  $$\sqrt(T_w)\sim 1/\sqrt{N}$$.\\
Another zone in the plot is located at the far right of the figure where $T_w/T_{eddy}$ is an $O(1)$ quantity so that the number of windows ($N_w$) is small for a specified simulation length ($T$). Undefined behavior of this zone is due to the invalidity of Equation (\ref{equ:SEM}) as the law of large numbers is not satisfied. The middle zone, where the figure lines are flat and collapsed, is the desired zone, where SEM is independent of the choice of $T_w$. This zone is where both $T_w$ and $N$ are sufficiently large to meet the (lack of) correlatiuon and law of large numbers requirements. However, this flat zone does not appear unless $T$ is sufficiently large. For HIT and for both quantities of interest,  ${u^\prime}^2$ and $S_{ij}S_{ij}$ , the flat zone appears when $T/T_{eddy} \ge 16000$, indicating a stringent simulation duration requirement for reliable reporting of SEM.\\

The intersection of straight lines that fit the asymptotic behavior of the highly correlated zone to left of the figure and the flat zone in the middle provides an estimate for minimum size for $T_w$. For our single point, second order statistics (${u^\prime}^2$ and $S_{ij}S_{ij}$), intersection happens around $T_w/T_{eddy} = O(10)$ for both simulated Reynolds numbers for this specific forcing scenario.

We extended our study to shear flows by simulating turbulent channel flow at $Re_\tau=180$ inside the channel with dimension of $2\pi, 1, \pi$. In this case we picked stream-wise velocity as the quantity of interest to calculate statistical error of the centerline mean velocity. For these simulations we adopted the simulation code originally developed by \cite{Bose} and then modified by \cite{Seo}. The fluids equations for channel flow are solved using a rectangular uniform mesh in $x$ and $z$ and non-uniform mesh in $y$ following a $tanh$ mapping function \cite{Bose}. The velocity data are located at a staggered configuration relative to the pressure data, and the spatial derivatives are discretized using the standard second-order finite difference scheme. The second-order Adams-Bashforth \cite{Morinishi} scheme is adopted for time stepping. 

Figure \ref{fig:Channel} shows that in order to achieve a flat zone in the plots a long simulation $(T/T_{eddy} \approx 14000)$ is required, which is similar in order of magnitude to the time scale revealed from the homogeneous turbulence analysis. Likewise, the intersection of asymptotes to the highly correlated zone and the middle zone indicates a time scale of $O(10)$ eddy turnover times. In this case $T_{eddy}$ is defiend based on the shear velocity and channel half height. 

Based on these observations, we conclude with the answers to the questions posed in this study. The requirement for uncorrelated samples in using the SEM equations (Equation~\ref{equ:SEM}), practically translates to sampling windows with duration larger than 10 eddy turnover times. This estimate is based on intersection of asymptotic lines modeling the behavior of highly correlated zones and uncorrelated zones in Figures 2 and 3. However, the actual behavior involves a transition between these asymptotic limits, with a gradual decline of correlation effects. Secondly, reliable reporting of SEM requires overall simulation times on the order of $10^4$ eddy turnover times. Once these criteria are met the central limit theorem can be naturally applied to offer reliable confidence intervals in reporting the estimated means from simulation data. A typical convention is to report $95\%$ confidence interval by multiplying SEM with 1.96 \cite{Ross}. This process assumes a Gaussian distribution of the error of the mean, which follows from the central limit theorem, given the error of the mean is a result of superposition of errors from uncorrelated windows. 

All simulations reported in the literature that we are aware of fall short in meeting the second criteria quantified by our analysis. This implies unreliable reporting of error bounds around the means from these simulations. In such cases, only the order of magnitude of the error can be trusted. Verification of codes based on simulation-to-simulation comparison and specifically based on establishing overlaps in confidence intervals will be problematic in such cases.

\begin{figure}
    \centering
    \includegraphics[width=0.5\textwidth]{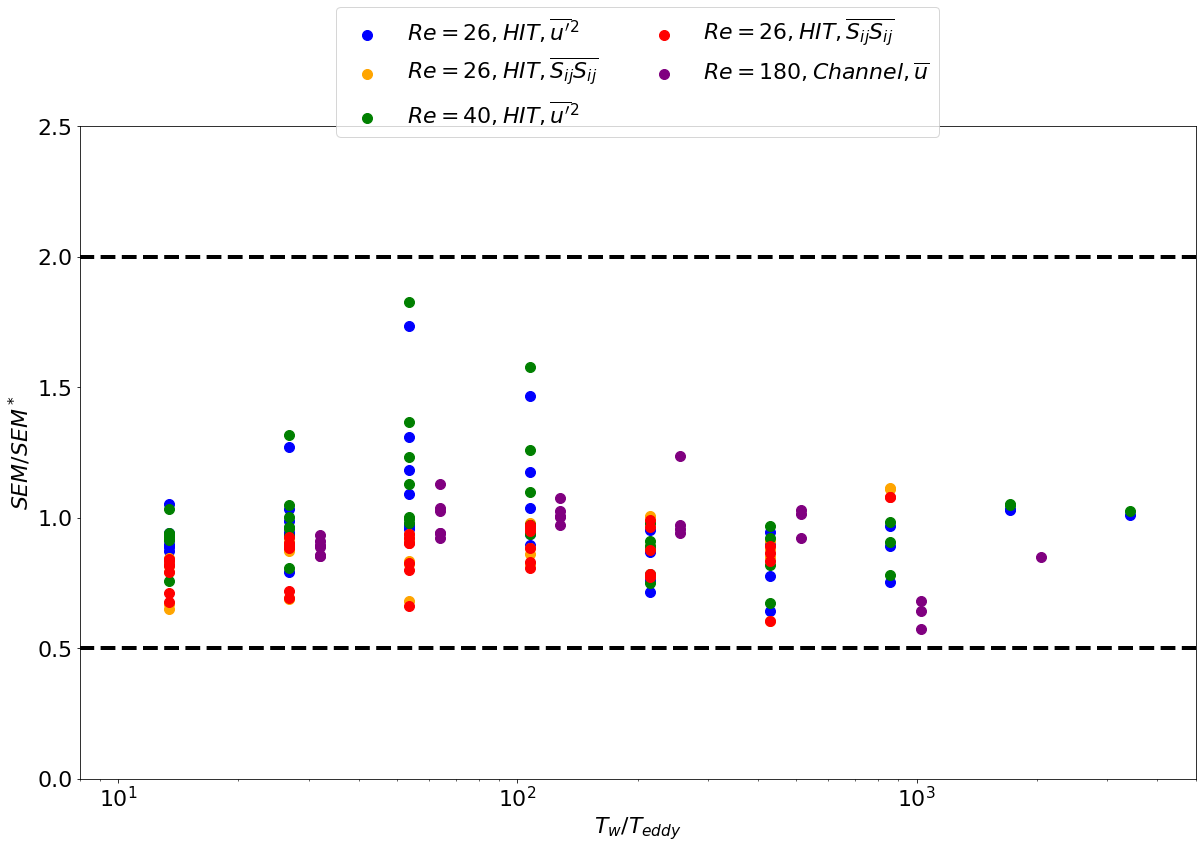}
    \caption{Normalized SEM for different simulations with choice of number of windows outside of the correlated data.}
    \label{fig:scatter}
\end{figure}


Lastly, we provide an estimate of error in determining confidence intervals when a simulation duration is not sufficiently long to establish a flat zone in estimation of SEM. In this case, instead of establishing a flat zone, we consider a weaker constraint given as $T_w> 10 T_{eddy}$ and $N\le 4$. Figure 4 shows computed SEM data from all simulations considered in this study in which these two criteria are met. For each case SEM is normalized by the asymptotic SEM from the flat zone in each figure. The data scatter indicates that the true SEM is within factor of 2, marked by dashed lines, compared to the computed SEM. In other words, in the absence of flat zones in SEM plots, which is the case for most practical simulations, one should consider roughly a factor of 2 error in their estimation of confidence intervals. This observation may explain mismatch between simulations and experiments when only nominally computed SEM is used for estimation of confidence intervals.

\section{Acknowledgments}

This study was supported by the U.S. Department of Energy, under grant number DE-NA0002373. This work was performed under the auspices of the U.S. Department of Energy by Lawrence Livermore National Laboratory under contract DE-AC52-07NA27344. Lawrence Livermore National Security, LLC. LLNL-JRNL-822749. 

\providecommand{\noopsort}[1]{}\providecommand{\singleletter}[1]{#1}%

\end{document}